\documentclass[twocolumn,showpacs,preprintnumbers,amsmath,amssymb]{revtex4}
\usepackage{graphicx}
\usepackage{dcolumn}
\usepackage{bm}

\begin{document}

\title{Solving real time evolution problems by constructing excitation operators}
\author{Pei Wang}
\email{pei.wang@live.com}
\affiliation{Institute of applied physics, Zhejiang University of Technology, Hangzhou, P. R. China}

\date{\today}

\begin{abstract}
In this paper we study the time evolution of an observable in the interacting fermion systems driven out of equilibrium. We present a method for solving the Heisenberg equations of motion by constructing excitation operators which are defined as the operators $\hat A$ satisfying $[\hat H,\hat A]=\lambda \hat A$. It is demonstrated how an excitation operator and its excitation energy $\lambda$ can be calculated. By an appropriate supposition of the form of $\hat A$ we turn the problem into the one of diagonalizing a series of matrices whose dimension depends linearly on the size of the system. We perform this method to calculate the evolution of the creation operator in a toy model Hamiltonian which is inspired by the Hubbard model and the nonequilibrium current through the single impurity Anderson model. This method is beyond the traditional perturbation theory in Keldysh-Green's function formalism, because the excitation energy $\lambda$ is modified by the interaction and it will appear in the exponent in the function of time. 
\end{abstract}


\maketitle

\section{Introduction}

The interacting fermion systems have a central position in modern condensed matter physics~\cite{schneider}. Their properties have been intensively studied both in theory and experiment. Good agreements between the theoretical predictions and the experimental results have been obtained when these systems are in thermal equilibrium. Examples include the Coulomb blockade effect and the Kondo effect in quantum dots~\cite{anderson61,kondo,hewson,meir,jauho,reed,wilson75,bulla}, and the charge density wave in the low-dimensional materials, e.g. the carbon nanotubes and the edge states in fractional quantum Hall effects~\cite{tomonaga50,luttinger63,wen90,kane,egger,delft}. However, how to understand the real time evolution of an interacting system driven out of equilibrium remains a great challenge in spite of intense efforts in recent years. The standard tool for the quantum field theory of nonequilibrium states is the Keldysh-Green's function technique. But the Keldysh techniques suffer from several shortcomings. First it is based on the Wick's theorem, and then the initial states must be limited to the equilibrium states of the quadratic Hamiltonian. Secondly, the result from a perturbative expansion with respect to the interaction strength does not incorporate the physics of the problem when the Coulomb interaction is the largest energy scale. Finally, for some models, e.g., the Kondo model, the perturbation theory is plagued by the infrared divergence at low temperatures.

For these reasons, new theoretical tools have been developed to understand the interacting fermion systems in nonequilibrium. Numerical methods are developed like the real time Quantum Monte Carlo~\cite{schiro09,werner09,schmidt08,muehlbacher08}, the time-dependent numerical renormalization group~\cite{costi,anders05,anders06}, the scattering state numerical renormalization group~\cite{anders08}, the adaptive time-dependent density matrix renormalization group~\cite{white04,daley04,vidal03} and the non-equilibrium dynamical mean field theory~\cite{schmidt02}. The numerical method has the advantage that the result can be obtained in very high precision and the parameters of the model can be chosen arbitrarily. However, they also suffer the inconvenience that the large computer resources are required and the result does not incorporate the physics of the problem directly.

The analytical methods are developed, which can supplement the shortcomings of the numerical methods, such as the scattering Bethe-ansatz~\cite{mehta}, the method based on integrability~\cite{schiller,lesage}, the real-time renormalization group~\cite{schoeller00,schoeller09,karrasch,pletyukhov}, the nonequilibrium flow equation~\cite{kehreinbook,hackl07,hackl08,eckstein,moeckel08}, and various approximation schemes building on the Green's function techniques~\cite{plihal,spataru}. However, no unique method is available which can cover all regimes of interest in different models. It is still necessary to study new methods.

The Hamiltonian of a typical interacting fermion system can be expressed as
\begin{eqnarray}\label{hamiltonian}\nonumber
\hat H= \sum_k \epsilon_k :\hat c^\dag_k \hat c_k: + && \sum_{k'_1,k'_2,k_1,k_2}U_{k'_1 k'_2 k_1 k_2} \\ && \times :\hat c^\dag_{k'_1} \hat c^\dag_{k'_2} \hat c_{k_2}\hat c_{k_1} :,
\end{eqnarray}
where $\hat c_k$ ($\hat c^\dag_k$) is the fermionic annihilation (creation) operator. And the normal ordering is with respect to the non-interacting ground state of the Fermi sea. In the traditional diagrammatic expansion the time evolution of a physical observable, denoted as $\langle \hat O(t) \rangle$ where $\hat O(t)$ satisfies the Heisenberg equation of motion
\begin{eqnarray}
 i \frac{d}{dt} \hat O(t)= [\hat O(t),\hat H],
\end{eqnarray}
is calculated by expanding the S-matrix and summing up a series of diagrams in a self-consistent way. In this paper we propose an alternate way to solve the Heisenberg equation of motion by decomposing the observable operator $\hat O$ into the linear combination of the excitation operators which are defined as the operator $\hat A$ satisfying the eigen equation
\begin{eqnarray}
 [\hat H, \hat A]=\lambda \hat A.
\end{eqnarray}
This method circumvents the expansion of the S-matrix and then the contraction of field operators, and is applicable even the initial state is nontrivial. Furthermore, as we will show, this method is beyond the traditional perturbation theory because the excitation energy $\lambda$ which appears in the exponent in the time function is modified by the interaction strength.

In Sec.~II we give the definition of the excitation operators. In Sec.~III we demonstrate how to construct the excitation operator and use it to calculate the time evolution of the creation operator in a toy model inspired by the Hubbard model. And the comparison between this model and the nonequilibrium flow equation approach which also contributes to solve the Heisenberg equation of motion is given. In Sec.~IV we perform our method in single impurity Anderson model and calculate the evolution of the current operator. Sec.~V is a short summary.

\section{excitation operators}

The excitation operator~\cite{fan} of a Hamiltonian $\hat H$ is the operator $\hat A$ which satisfies
\begin{eqnarray}\label{eigenequation}
[\hat H, \hat A] =\lambda \hat A, 
\end{eqnarray}
where $\lambda$ is a real number denoting the excitation energy. We suppose that $| \psi_x\rangle$ and $| \psi_y\rangle$ are two arbitrary eigen-states of $\hat H$ with the eigen-energies $E_x$ and $E_y$ respectively. Then the excitation operator can be expressed as 
\begin{eqnarray}\label{excitation_eigenstates}
\hat A_{x,y} = | \psi_x\rangle \langle \psi_y |. 
\end{eqnarray}
The corresponding excitation energy is $\lambda_{x,y}=E_x-E_y$.
An observable operator $\hat O$ can be decomposed into the linear combination of the excitation operators, i.e.,
\begin{eqnarray}
\hat O = \sum_{x,y} O_{x,y} \hat A_{x,y} ,
\end{eqnarray} 
where $O_{x,y}=\langle \psi_x|\hat O|\psi_y\rangle$. According to Eq.~\ref{eigenequation}, we have
\begin{eqnarray}\label{aevolution}
e^{i\hat Ht}\hat A_{x,y} e^{-i\hat Ht} = e^{i\lambda_{x,y} t} \hat A_{x,y}.
\end{eqnarray}
The solution of the Heisenberg equation of motion can be expressed as
\begin{eqnarray}\label{evolution}
\hat O (t) = \sum_{x,y} O_{x,y} e^{i\lambda_{x,y} t} \hat A_{x,y}.
\end{eqnarray}
Once we know how to decompose an observable operator, we obtain the time evolution of it. 

In many-body problems it is generally difficult to find the eigen-states of the Hamiltonian and then calculate the excitation operator according to Eq.~\ref{excitation_eigenstates}. An alternate way is to expand $\hat A$ into a series of products of field operators, substitute the expansion into Eq.~\ref{eigenequation} and decide the coefficients. For the Hamiltonian of Eq.~\ref{hamiltonian}, we could express $\hat A$ as
\begin{eqnarray}\nonumber\label{expansion_power}
\hat A &=& \sum_k M_k \hat c^\dag_k + \sum_{k,k'} M_{k'k} :\hat c^\dag_{k'} \hat c_k: \\ && + \sum_{k'_1k'_2k_1} M_{k'_1k'_2k_1} :\hat c^\dag_{k'_1} \hat c^\dag_{k'_2} \hat c_{k_1}:+ \cdots,
\end{eqnarray}
where $M_k$, $M_{k'k}$ and $M_{k'_1k'_2k_1}$ are the undetermined coefficients. The commutator of the Hamiltonian with the products of $N$ field operators contains the products of $(N+2)$ operators, which indicate that $\hat A$ should be an infinite series. We need to truncate the series to get an approximated expression of $\hat A$.

Since we employ the perturbative expansion in the expression of $\hat A$, our method is also a kind of perturbative approach. While the direct perturbative approach for solving the Heisenberg equations of motion will generate secular terms, which are uncontrolled at large timescales~\cite{hackl07,hackl08}. By first decomposing the observable into the linear combination of the excitation operators, we avoid the secular divergences and make the results about the long-time behavior more explicit.

\section{a toy model Hamiltonian with degenerate interactions}
\subsection{Excitation operators}

In this section, we use a toy model Hamiltonian
\begin{eqnarray}\nonumber\label{modelhamiltonian}
\hat H &=& \sum_{k,\sigma} \epsilon_k :\hat c^\dag_{k\sigma}\hat c_{k\sigma}: +\frac{U}{2}\sum_{k\neq k',\sigma}:\hat c^\dag_{k'\sigma}\hat c^\dag_{k\bar\sigma}\hat c_{k\bar \sigma}\hat c_{k'\sigma} : \\ && + \frac{U}{2}\sum_{k\neq k',\sigma}:\hat c^\dag_{k'\sigma}\hat c^\dag_{k\bar\sigma} \hat c_{k'\bar\sigma} \hat c_{k\sigma}:
\end{eqnarray}
to demonstrate how to construct excitation operators and use them to calculate the evolution of an observable operator. In Eq.~\ref{modelhamiltonian} the first term denotes the single particle energy, and the second and third terms the interactions. Here $\sigma=\uparrow,\downarrow$ denotes the particle spin, and $\bar \sigma$ the opposite to $\sigma$. The normal ordering is with respect to the Fermi sea with chemical potential $\mu$. The Fermi distribution function is denoted as $n_k=\theta(\mu-\epsilon_k)$. It is not difficult to find that this model Hamiltonian is inspired by the Hubbard model. In fact it is the simplified Hubbard model in which most of the interaction terms are thrown off. Only the degenerate interaction term is left which is defined as the interaction keeping the sum of the single particle energies of the two annihilation operators exactly as same as that of the two creation operators. For example, the degenerate interaction in the Hamiltonian of Eq.~\ref{hamiltonian} should be the interaction terms $U_{k'_1 k'_2 k_1 k_2}:\hat c^\dag_{k'_1} \hat c^\dag_{k'_2} \hat c_{k_2}\hat c_{k_1} :$ in which the relation $\epsilon_{k_1}+\epsilon_{k_2}=\epsilon_{k'_1}+\epsilon_{k'_2}$ is satisfied.

We are interested in calculating $e^{i\hat H t} \hat c^\dag_{k\sigma} e^{-i \hat H t}$. We notice that $\hat c^\dag_{k\sigma}$ increases the particle number of a state by one. Then it will be decomposed into the excitation operators in which arbitrary term contains one more creation operator than the annihilation operator. We use the symbol $\hat A_{k\sigma}$ to represent the excitation operators. Again we could suppose that the first term of $\hat A_{k\sigma}$ is $\hat c^\dag_{k\sigma}$ in the simplest case. The commutator of $\hat H$ with $\hat c^\dag_{k\sigma}$ contains the terms $: \hat c^\dag_{k\sigma}\hat c^\dag_{k'\bar \sigma}\hat c_{k'\bar \sigma}:$ and $:\hat c^\dag_{k'\sigma}\hat c^\dag_{k\bar \sigma} \hat c_{k'\bar \sigma}:$. The commutator of $\hat H$ with $: \hat c^\dag_{k\sigma}\hat c^\dag_{k'\bar \sigma}\hat c_{k'\bar \sigma}:$ and $:\hat c^\dag_{k'\sigma}\hat c^\dag_{k\bar \sigma} \hat c_{k'\bar \sigma}:$ contains one more term $:\hat c^\dag_{k\sigma}\hat c^\dag_{k'\sigma}\hat c_{k'\sigma}:$. So we could suppose the expression of the excitation operator as
\begin{eqnarray}\label{aexpression_normal}\nonumber
\hat A_{k \sigma} &=& \hat c^\dag_{k\sigma}+ \sum_{k' \neq k} (M^{1}_{k,k'}: \hat c^\dag_{k\sigma}\hat c^\dag_{k'\bar \sigma}\hat c_{k'\bar \sigma}: \\ && \nonumber+ M^{2}_{k,k'}:\hat c^\dag_{k'\sigma}\hat c^\dag_{k\bar \sigma} \hat c_{k'\bar \sigma}: + M^{3}_{k,k'}:\hat c^\dag_{k\sigma}\hat c^\dag_{k'\sigma}\hat c_{k'\sigma}:)\\ && +\cdots .
\end{eqnarray}
In Eq.~\ref{aexpression_normal} we neglect the high order terms consisting of more than three field operators. 

The eigen-equation is written as 
\begin{eqnarray}
[\hat H,\hat A_{k\sigma}]= \lambda_k \hat A_{k\sigma}, 
\end{eqnarray}
where $\lambda_k$ is the excitation energy. Substituting Eq.~\ref{modelhamiltonian} and \ref{aexpression_normal} in and comparing the coefficients before $\hat c^\dag_{k\sigma}$ between the left and right sides of the equation, we get
\begin{eqnarray}
\lambda_k = \epsilon_k +  U\sum_{k' \neq k} (M^1_{k,k'}+M^2_{k,k'})(-n^2_{k'}+n_{k'}) .
\end{eqnarray}
But $-n^2_{k'}+n_{k'} = 0$ whether $n_{k'}=0$ or $n_{k'}=1$. Then the excitation energy becomes
\begin{eqnarray}
\lambda_k= \epsilon_k.
\end{eqnarray}
Comparing the coefficients before $: \hat c^\dag_{k\sigma}\hat c^\dag_{k'\bar \sigma}\hat c_{k'\bar\sigma}:$, $:\hat c^\dag_{k'\sigma}\hat c^\dag_{k\bar \sigma} \hat c_{k'\bar\sigma}:$ and $:\hat c^\dag_{k\sigma}\hat c^\dag_{k'\sigma}\hat c_{k'\sigma}:$ between the two sides of the eigen-equation for $k\neq k'$, we get
\begin{equation}\label{equationMtoy}
\begin{split}
(\lambda_k-\epsilon_k) M^1_{k,k'} =& U \left( 1+M^1_{k,k'} (1-2n_{k'}) \right. \\ & \left. + M^2_{k,k'}(1-n_k-n_{k'}) \right), \\ 
(\lambda_k-\epsilon_k) M^2_{k,k'} =& U \left( 1+M^1_{k,k'}(1-n_k-n_{k'}) \right. \\ & \left. + M^2_{k,k'}(1-n_k-n_{k'}) \right. \\ & \left. + M^3_{k,k'} (n_k-n_{k'}) \right), \\ 
(\lambda_k-\epsilon_k) M^3_{k,k'} =& U M^2_{k,k'} (n_k-n_{k'}).
\end{split}
\end{equation}
These equations are solved and the coefficients $M$ are decided. When $n_k=n_{k'}$ we find that 
\begin{eqnarray}\label{mequation_normal}
M^1_{k,k'} + M^2_{k,k'}= \frac{1}{2n_{k'}-1},
\end{eqnarray}
and $M^3_{k,k'}$ is an arbitrary number. What we want is a set of linearly-independent excitation operators. So we set $M^3_{k,k'}=0$, since $:\hat c^\dag_{k\sigma} \hat c^\dag_{k'\sigma} \hat c_{k'\sigma}: $ is itself an excitation operator as will be shown next. Eq.~\ref{mequation_normal} has two linearly-independent solutions: $M^1_{k,k'}=0$, $M^2_{k,k'}= \frac{1}{2n_{k'}-1}$; and $M^1_{k,k'}= \frac{1}{2n_{k'}-1}$, $M^2_{k,k'}=0$. One could choose arbitrary one for $\hat A_{k\sigma}$. We use an indicator function $\chi_{k,k'}=0,1$ to denote our choice, and express the corresponding term in $\hat A_{k\sigma}$ as
\begin{equation}
 \begin{split}
& \frac{1}{2n_{k'}-1} \left( \chi_{k,k'} :\hat c^\dag_{k\sigma}\hat c^\dag_{k'\bar \sigma}\hat c_{k'\bar\sigma}: \right. \\ & \left. + (1-\chi_{k,k'}):\hat c^\dag_{k'\sigma}\hat c^\dag_{k\bar \sigma}\hat c_{k'\bar\sigma}: \right).
\end{split}
\end{equation}

For $n_k \neq n_{k'}$, the solution of Eq.~\ref{equationMtoy} is
\begin{eqnarray}\nonumber
M^1_{k,k'} &=& \frac{1}{2n_{k'}-1}, \\ \nonumber
M^2_{k,k'} &=& 0, \\ 
M^3_{k,k'} &=& \frac{1}{n_{k'}-n_k}.
\end{eqnarray}
In summary the excitation operator can be expressed as 
\begin{equation}
\begin{split}
\hat A_{k \sigma} =& \hat c^\dag_{k\sigma}+ \sum_{n_{k'}\neq n_k} \left( \frac{1}{2n_{k'}-1} :\hat c^\dag_{k\sigma}\hat c^\dag_{k'\bar \sigma}\hat c_{k'\bar\sigma}: \right. \\ & \left. +\frac{1}{n_{k'}-n_k} :\hat c^\dag_{k\sigma}\hat c^\dag_{k'\sigma}\hat c_{k'\sigma}: \right) \\  & +  \sum_{k'\neq k,n_{k'}=n_k}\frac{1}{2n_{k'}-1} \left(\chi_{k,k'} :\hat c^\dag_{k\sigma}\hat c^\dag_{k'\bar \sigma}\hat c_{k'\bar\sigma}: \right. \\ & \left. +(1-\chi_{k,k'}):\hat c^\dag_{k'\sigma}\hat c^\dag_{k\bar \sigma}\hat c_{k'\bar\sigma}: \right) .
\end{split}
\end{equation}
Different indicator functions will give different $\hat A_{k \sigma}$. But we will show next that the choice of $\chi_{k,k'}$ does not affect the last result of $e^{i\hat H t} \hat c^\dag_{k\sigma} e^{-i \hat H t}$.

To decompose $\hat c^\dag_{k\sigma}$ we need to cancel the terms $:c^\dag_{k\sigma}c^\dag_{k'\bar \sigma} c_{k'\bar\sigma}:$, $:\hat c^\dag_{k'\sigma}\hat c^\dag_{k\bar \sigma}\hat c_{k'\bar\sigma}:$ and $:\hat c^\dag_{k\sigma}\hat c^\dag_{k'\sigma}\hat c_{k'\sigma}:$ in $\hat A_{k \sigma}$. So we construct the excitation operators with them as the leading terms. To avoid ambiguity we use the symbol $\hat B$ to denote these operators. For arbitrary $k\neq k'$ we suppose that
\begin{eqnarray}\nonumber
\hat B_{kk' \sigma} &=& N^{1}_{k,k'} :\hat c^\dag_{k\sigma}\hat c^\dag_{k'\bar \sigma}\hat c_{k'\bar\sigma}: + N^{2}_{k,k'}:\hat c^\dag_{k'\sigma}\hat c^\dag_{k\bar \sigma}\hat c_{k'\bar\sigma}: \\ && + N^{3}_{k,k'}:\hat c^\dag_{k\sigma} \hat c^\dag_{k'\sigma} c_{k'\sigma}: ,
\end{eqnarray}
where we again neglect the high order terms consisting of more than three field operators. 
This supposition is reasonable because the commutator $[\hat H,\hat B_{kk' \sigma}]$ does not contain the terms like single creation operator. The eigen-equation is now written as 
\begin{eqnarray}
[\hat H,\hat B_{kk' \sigma}]=\lambda^B_{k,k'} \hat B_{kk' \sigma}.
\end{eqnarray}
By comparing the coefficients between the left and right sides of the equation we have
\begin{equation}
\begin{split}\label{eigenequation_N}
(\lambda^B_{k,k'} -\epsilon_k) N^1_{k,k'} =& U \left( N^1_{k,k'} (1-2n_{k'}) \right. \\ & \left. + N^2_{k,k'}(1-n_k-n_{k'}) \right) , \\
(\lambda^B_{k,k'}-\epsilon_k) N^2_{k,k'} =& U \left( (N^1_{k,k'}+N^2_{k,k'})(1-n_k-n_{k'}) \right. \\ & \left.  + N^3_{k,k'} (n_k-n_{k'}) \right) , \\ 
(\lambda^B_{k,k'}-\epsilon_k) N^3_{k,k'} =& U N^2_{k,k'} (n_k-n_{k'}).
\end{split}
\end{equation}
In the matrix form of this equation we find that $(N^1_{k,k'},N^2_{k,k'},N^3_{k,k'})$ is in fact the eigenvector of a $(3\times 3)$ matrix. When $n_k=n_{k'}$ Eq.~\ref{eigenequation_N} can be expressed in the matrix form as
\begin{eqnarray}\nonumber
&& \left ( \begin{array}{ccc} \epsilon_k + U(1-2n_{k'}) & U(1-2n_{k'})  & 0\\
U(1-2n_{k'}) & \epsilon_k + U(1-2n_{k'}) & 0 \\ 0 & 0 & \epsilon_k \end{array} \right ) \\ && \times \left (\begin{array}{c} N^1_{k,k'} \\ 
N^2_{k,k'} \\ N^3_{k,k'} \end{array} \right ) = \lambda^B_{k,k'} \left (\begin{array}{c} N^1_{k,k'} \\ 
N^2_{k,k'}  \\ N^3_{k,k'}  \end{array}\right ) .
\end{eqnarray}
This matrix has three linearly-independent eigenvectors. Correspondingly we get three linearly-independent $\hat B_{kk'\sigma}$. When $n_k\neq n_{k'}$ Eq.~\ref{eigenequation_N} becomes
\begin{eqnarray}\nonumber
&& \left ( \begin{array}{ccc} \epsilon_k + U(1-2n_{k'}) & 0 & 0 \\
0 & \epsilon_k & U(n_k-n_{k'})  \\ 0 & U(n_k-n_{k'}) & \epsilon_k \end{array} \right ) \\ && \times \left (\begin{array}{c} N^1_{k,k'} \\ N^2_{k,k'} \\ N^3_{k,k'} \end{array} \right ) = \lambda^B_{k,k'} \left (\begin{array}{c} N^1_{k,k'} \\ N^2_{k,k'} \\ N^3_{k,k'} \end{array} \right ),
\end{eqnarray}
where we have used the identity $n_k+n_{k'}=1$. This matrix contributes another three eigenvectors.
\begin{table}\label{table}
\caption{The six types of $\hat B_{kk'\sigma}$ operators}
\begin{center}
\begin{tabular}{ccc}
\hline \hline
Name & Expression & Energy \\
$\hat B^{1}$ & $:\hat c^\dag_{k\sigma} \hat c^\dag_{k'\bar \sigma} \hat c_{k'\bar\sigma}: 
+ :\hat c^\dag_{k'\sigma} \hat c^\dag_{k\bar \sigma} c_{k'\bar\sigma}:$  & $\epsilon_k+ 2U(1-2n_{k'})$ \\
$\hat B^{2}$ & $:\hat c^\dag_{k\sigma}\hat c^\dag_{k'\bar \sigma} c_{k'\bar\sigma}: 
-:\hat c^\dag_{k'\sigma} \hat c^\dag_{k\bar \sigma} \hat c_{k'\bar\sigma}:$  & $\epsilon_k $ \\
$\hat B^{3}$ & $:\hat c^\dag_{k\sigma}\hat c^\dag_{k' \sigma} \hat c_{k' \sigma}: $  & $\epsilon_k$ \\
$\hat B^{4}$ & $:\hat c^\dag_{k\sigma}\hat c^\dag_{k'\bar \sigma} \hat c_{k'\bar\sigma}: $  & $\epsilon_k+ U(1-2n_{k'})$ \\
$\hat B^{5}$ & $:\hat c^\dag_{k'\sigma} \hat c^\dag_{k\bar \sigma} \hat c_{k'\bar\sigma}: + :\hat c^\dag_{k\sigma}\hat c^\dag_{k' \sigma} \hat c_{k' \sigma}: $  & $\epsilon_k+ U(n_k-n_{k'})$ \\
$\hat B^{6}$ & $:\hat c^\dag_{k'\sigma} \hat c^\dag_{k\bar \sigma} \hat c_{k'\bar\sigma}: - :\hat c^\dag_{k\sigma}\hat c^\dag_{k' \sigma} \hat c_{k' \sigma}: $  & $\epsilon_k - U(n_k-n_{k'})$ \\
\hline
\end{tabular}
\end{center}
\end{table}
Totally we obtain six types of $\hat B_{kk'\sigma}$, which are all listed in table~I with the corresponding excitation energies. 

In summary we get the excitation operators $\hat A_{k\sigma}$ and $\hat B_{kk'\sigma}$. We could also get the excitation operators of higher orders by similar analysis. Even we only keep the lowest order terms in the perturbative expansion of $\hat B_{kk'\sigma}$, we get a nontrivial result. The interaction strength $U$ enters the expression of the excitation energy of $\hat B_{kk'\sigma}$. We conclude that $\hat B_{kk'\sigma}$ reflects the characteristics of the collective excitations. When $n_{k'}=0$ we find that $\hat B^1_{kk'\sigma} = (\hat c^\dag_{k\sigma}\hat c^\dag_{k'\bar\sigma}-  \hat c^\dag_{k\bar\sigma} \hat c^\dag_{k'\sigma} ) \hat c_{k'\bar \sigma}$ and the corresponding excitation energy is $\epsilon_k +2U$. This operator annihilates an electron and simultaneously creats a spin singlet. This procedure costs an extra energy of $2U$. 
  
\subsection{The evolution of single creation operator}

The operator $\hat c^\dag_{k\sigma}$ can be decomposed into the linear combination of the operators $\hat A_{k\sigma}$ and $\hat B_{kk' \sigma}$. The second order terms in $\hat A_{k\sigma}$ with $n_k=n_{k'}$ can be canceled by $\hat B^{1}_{kk'\sigma}$, $\hat B^{2}_{kk'\sigma}$ and $\hat B^{3}_{kk'\sigma}$, and those with $n_k\neq n_{k'}$ by $\hat B^{4}_{kk'\sigma}$, $\hat B^{5}_{kk'\sigma}$ and $\hat B^{6}_{kk'\sigma}$. It is easy to find
\begin{equation}
\begin{split}
\hat c^\dag_{k\sigma} =& \hat A_{k\sigma}-  \sum_{k'\neq k,n_{k'}=n_k}\frac{1}{2n_{k'}-1} \left(\frac{1}{2} \hat B^{1}_{kk'\sigma} \right. \\ & \left. +(\chi_{k,k'}-\frac{1}{2}) \hat B^{2}_{kk'\sigma} \right )\\ &  - \sum_{n_{k'}\neq n_k} \left(\frac{1}{2n_{k'}-1} \hat B^{4}_{kk'\sigma} +\frac{1}{2(n_{k'}-n_k)}\hat B^{5}_{kk'\sigma} \right.\\ & \left. -\frac{1}{2(n_{k'}-n_k)}\hat B^{6}_{kk'\sigma}  \right). 
\end{split}
\end{equation}
The expressions of $e^{i\hat Ht} \hat A_{k\sigma} e^{-i\hat Ht}$ and $e^{i\hat Ht}\hat B_{kk'\sigma}e^{-i\hat Ht}$ are got from Eq.~\ref{aevolution} and the excitation energies listed in table~I. Substituting back in the expression of $\hat A_{k\sigma}$ and $\hat B_{kk'\sigma}$ we get
\begin{equation}\label{evolutionresult}
 \begin{split}
\hat c^\dag_{k\sigma} (t) =& e^{i\epsilon_k t} \hat c^\dag_{k\sigma}+  \sum_{k'\neq k,n_{k'}=n_k}  \frac{e^{i\epsilon_k t}(1-e^{2it U(1-2n_{k'})}) }{2(2n_{k'}-1)}\\ & \times (:\hat c^\dag_{k\sigma}\hat c^\dag_{k'\bar \sigma} \hat c_{k'\bar\sigma}: 
+ :\hat c^\dag_{k'\sigma} \hat c^\dag_{k\bar \sigma} \hat c_{k'\bar\sigma}:) \\ &  + \sum_{n_{k'}\neq n_k} \left(\frac{e^{i\epsilon_k t}(1-e^{it U(1-2n_{k'})} )}{2n_{k'}-1} :\hat c^\dag_{k\sigma} \hat c^\dag_{k'\bar \sigma} \hat c_{k'\bar\sigma}: \right. \\ & \left. +\frac{ie^{it \epsilon_k} \sin [t U(n_{k'}-n_{k})] }{n_{k'}-n_k} :\hat c^\dag_{k'\sigma}\hat c^\dag_{k \bar\sigma}\hat c_{k' \bar \sigma}: \right. \\ & \left. +\frac{e^{it \epsilon_k}(1-\cos [t U(n_k-n_{k'})] )}{n_{k'}-n_k} :\hat c^\dag_{k\sigma}\hat c^\dag_{k'\sigma}\hat c_{k'\sigma}:   \right).
\end{split}
\end{equation}
Note that the expression of $e^{i\hat Ht}\hat c^\dag_{k\sigma} e^{-i\hat Ht}$ is unique, independent to the indicator function.

In Eq.~\ref{evolutionresult} the function of time is $U$-dependent, which indicates that our method involves the renormalization of the excitation energies and is beyond the traditional perturbation theory. This equation is an operator identity. As the operator $\hat c^\dag_{k\sigma} (t)$ acting on $e^{i\hat Ht}|\psi\rangle$ where $|\psi\rangle$ is an eigenstate, we get $e^{i\hat Ht}\hat c^\dag_{k\sigma}|\psi\rangle$. So the physical meaning of Eq.~\ref{evolutionresult} is that it shows how the system responses to a single particle excitation by producing the collective excitations.

It is worthwhile to mention that the nonequilibrium flow equation~\cite{hackl07,hackl08} is also a method designed for solving the Heisenberg equation of motion by expanding the operator into a power series. There are three main difference between our method and the flow equation approach. First in the flow equation approach a unitary transformation is performed to the Hamiltonian, while in our method the Hamiltonian keeps invariant. Secondly, we turns the problem into the diagonalization of the coefficient matrix instead of solving the differential equation in the flow equation approach. Finally we deal with the degenerate interaction in the same way as the non-degenerate ones, which suggests that this method is applicable in the model where the degenerate interaction is overwhelming, e.g. in the Tomonaga-Luttinger model~\cite{delft}. While in the flow equation approach the degenerate interaction has a different scaling behavior in the flow.

\section{single impurity Anderson model}

Next we consider the single impurity Anderson model, because it is important both in theory and experiment. Furthermore, many papers have contributed to study this model, so that we could compare our method with the others. We will use the excitation operators to calculate the time evolution of the current through the impurity. 

The Hamiltonian of the Anderson impurity model is expressed as
\begin{eqnarray} \nonumber\label{andersonimpurityH}
\hat H &=& \sum_{k\alpha\sigma} \epsilon_k \hat c^\dag_{k\alpha \sigma} \hat c_{k\alpha \sigma} + \sum_{k\alpha\sigma} \frac{V}{\sqrt{2}} (\hat c^\dag_{k \alpha \sigma} \hat d_\sigma + h.c.)\\ && - \frac{U}{2} \sum_\sigma\hat d^\dag_\sigma \hat d_\sigma + U \hat d^\dag_\uparrow \hat d_\uparrow \hat d^\dag_\downarrow \hat d_\downarrow, 
\end{eqnarray}
where $\sigma=\uparrow,\downarrow$ denotes the electron spin and $\alpha=L,R$ the left and right leads. Here $\hat c^\dag_{k\alpha \sigma}$ and $\hat d^\dag_\sigma$ are the single-electron creation operators in the leads and at the impurity site respectively. At initial time the leads are in thermal equilibrium with chemical potentials $\pm \frac{V_{sd}}{2}$ respectively. The coupling between leads and the impurity is switched on at time $t=0$. 

In the pre-diagonalization basis~\cite{moeckelthesis,pei10}, the Hamiltonian can be expressed as
\begin{eqnarray}\nonumber
\hat H &=&  \sum_{k\sigma} \epsilon_k \hat c^\dag_{k - \sigma} \hat c_{k - \sigma} + \sum_{s\sigma} \epsilon_s \hat c^\dag_{s \sigma} \hat c_{s \sigma}\\ &&
 + \sum_{s'_1s'_2s_1s_2} U  B_{s'_1s'_2s_1s_2} :\hat c^\dag_{s_1' \uparrow} \hat c^\dag_{s_2' \downarrow}\hat c_{s_2 \downarrow}\hat c_{s_1 \uparrow}:,
 \end{eqnarray}
where $B_{s'_1s'_2s_1s_2}$ is the abbreviation of $B_{s_1'}B_{s_1}B_{s_2'}B_{s_2}$ and the normal ordering is with respect to the initial state. The (anti)symmetric operator is defined as $\hat c_{k \pm \sigma}=\frac{1}{\sqrt{2}} (\hat c_{k L \sigma} \pm \hat c_{k R \sigma})$ and the hybridization operator as $\hat c_{s \sigma} = \sum_k \frac{V}{\epsilon_s - \epsilon_k} B_s \hat c_{k + \sigma} + B_s \hat d_\sigma$. Here the coefficient $B_s = \frac{V}{\sqrt{\epsilon_s^2 + \Gamma ^ 2}}$ and the linewidth $\Gamma=\rho \pi V^2$, where $\rho$ is the density of states of the lead. In the Anderson impurity model we will neglect the degenerate interaction by ordering $B_{s'_1s'_2s_1s_2}=0$ as $\epsilon_{s'_1}+\epsilon_{s'_2}=\epsilon_{s_1}+\epsilon_{s_2}$. This is reasonable in the Anderson impurity model, since the coefficient $B_s\to 0$ in the thermodynamic limit. The Fermi function in hybridization basis has sharp edges in the thermodynamic limit, i.e., 
\begin{eqnarray}
\langle c^\dag_{s'\sigma}c_{s\sigma} \rangle_0 = \delta_{s,s'}n_s,
\end{eqnarray}
where $n_s=\frac{1}{2}(\theta(\frac{V_{sd}}{2}-\epsilon_s)+\theta(-\frac{V_{sd}}{2}-\epsilon_s))$.

We are interested in the evolution of the current operator
\begin{eqnarray}\nonumber
 I_\uparrow && =\frac{1}{2}(\frac{d \hat N_{L \uparrow}}{dt} - \frac{d \hat N_{R \uparrow}}{dt} ) \\
&& =\frac{iV}{2} \sum_{s,k} B_s \hat c^\dag_ {s \uparrow}\hat c_{k - \uparrow} + h.c.,
\label{defcurrentop}
\end{eqnarray}
where the antisymmetric field operator $\hat c_{k-\uparrow}$ is just the excitation operator of the Hamiltonian and its evolution can be written as $e^{i\hat H t}\hat c_{k-\uparrow} e^{-i\hat H t}= e^{-i\epsilon_k t}\hat c_{k-\uparrow} $. So the point is how to calculate $e^{i\hat H t} \hat c^\dag_{s\uparrow}e^{-i\hat H t}$, which needs us to decompose $\hat c^\dag_{s\uparrow}$ into the excitation operators.

Because in the Hamiltonian~(\ref{andersonimpurityH}) the total electron number and spin are conserved, we suppose the excitation operator as
\begin{eqnarray}\nonumber\label{andersona}
 \hat A &=& \sum_s M^1_s \hat c^\dag_{s\uparrow} + \sum_{s'_1s'_2s_2} M_{s'_1s'_2s_2}^2 :\hat c^\dag_{s'_1\uparrow} \hat c^\dag_{s'_2\downarrow} \hat c_{s_2\downarrow}: \\ && + \sum_{s'_1s'_2s_2} M_{s'_1s'_2s_2}^3 :\hat c^\dag_{s'_1\uparrow} \hat c^\dag_{s'_2\uparrow} \hat c_{s_2\uparrow}:. 
\end{eqnarray}
Here we truncate the series of $\hat A$ and neglect the higher order terms consisting of more than three field operators. In this approximation we can obtain the current to order $U^2$. 

Substituting Eq.~\ref{andersona} into Eq.~\ref{eigenequation} and comparing the coefficients between the terms at the left and right sides of the equation, we get
\begin{eqnarray}\label{solvingmmatrix}\nonumber
&& \epsilon_s M^1_s + U \sum_{s'_1s'_2s_2} B_{ss'_1s'_2s_2}  Q_{s'_1s'_2s_2} M^2_{s'_1s'_2s_2} = \lambda M^1_s, \\
&& \nonumber (\epsilon_{s'_1}+\epsilon_{s'_2}-\epsilon_{s_2}) M^2_{s'_1s'_2s_2} + U\sum_s B_{s'_1s'_2s_2s}M^1_s \\ && \nonumber + U\sum_{t_1t_2}B_{s'_1 t_1s_2t_2} (n_{t_1}-n_{t_2}) M^2_{t_1s'_2t_2} \\&& \nonumber + U\sum_{t_1t_2} B_{t_1t_2s'_1s'_2}(1-n_{t_1}-n_{t_2}) M^2_{t_1t_2s_2}\\ && \nonumber + U \sum_{t_1t_2}B_{t_1t_2s'_2s_2} (n_{t_1}-n_{t_2})M^3_{t_1s'_1t_2} \\ && \nonumber + U \sum_{t_1t_2} B_{t_1t_2 s'_2s_2} (n_{t_2}-n_{t_1}) M^3_{s'_1t_1t_2} = \lambda M^2_{s'_1s'_2s_2}, \\ && \nonumber (\epsilon_{s'_1}+\epsilon_{s'_2}-\epsilon_{s_2})M^3_{s'_1s'_2s_2} + U\sum_{t_1t_2} B_{s'_2s_2 t_1t_2}(n_{t_1}-n_{t_2}) \\ && \times M^2_{s'_1 t_2t_1} = \lambda M^3_{s'_1s'_2s_2}, 
\end{eqnarray}
where $Q_{s'_1s'_2s_2}=n_{s'_1}n_{s'_2}-n_{s'_1}n_{s_2}-n_{s'_2}n_{s_2}+n_{s_2}$. We express above equations in a matrix form
\begin{eqnarray}
 \mathcal{H} \mathcal{M} = \lambda \mathcal{M},
\end{eqnarray}
where $\mathcal{M}=(M^1_s, M^2_{s'_1s'_2s_2}, M^3_{s'_1s'_2s_2})$ is a vector. The dimension of the vector $\mathcal{M}$ is $(N+2N^3)$, and it is the eigenvector of a $(N+2N^3)\times (N+2N^3)$ matrix $\mathcal{H}$, where $N$ is the number of possible values of $s$, i.e., the number of levels in each lead. And $\lambda$ is the corresponding eigenvalue. To find $(M^1_s, M^2_{s'_1s'_2s_2}, M^3_{s'_1s'_2s_2})$ satisfying Eq.~\ref{solvingmmatrix} is equivalent to diagonalize the matrix $\mathcal{H}$. Now the problem changes into the problem of diagonalizing a matrix whose dimension is much less than the dimension of the original Hamiltonian. The diagonal elements of $\mathcal{H}$ are $U$-independent, while the non-diagonal elements are in order $U$. When $U$ is small, we could employ perturbation theory to calculate the eigenvalues and the eigenvectors of $\mathcal{H}$. By setting $U=0$, we get the eigenvectors and the corresponding eigenvalues in zeroth order. There are three types of eigenvectors. They are 
\begin{eqnarray}
\mathcal{M}^1_s \sim (M^1_s=1, M^2_{s'_1s'_2s_2}=0, M^3_{s'_1s'_2s_2}=0)
\end{eqnarray}
with the eigenvalue $\epsilon_s$ and 
\begin{eqnarray}\nonumber
\mathcal{M}^2_{s'_1s'_2s_2} & \sim & (M^1_s=0, M^2_{s'_1s'_2s_2}=1, M^3_{s'_1s'_2s_2}=0), \\
\mathcal{M}^3_{s'_1s'_2s_2} & \sim & (M^1_s=0, M^2_{s'_1s'_2s_2}=0, M^3_{s'_1s'_2s_2}=1)
\end{eqnarray}
with the eigenvalue $\epsilon_{s'_1}+\epsilon_{s'_2}-\epsilon_{s_2}$. For simplicity, we express the components of the eigenvector $\mathcal{M}^1_s$ as $\mathcal{M}^{1,1}_{s,s'}=\delta_{s,s'}$, $\mathcal{M}^{1,2}_{s,s'_1s'_2s_2}=\mathcal{M}^{1,3}_{s,s'_1s'_2s_2}=0$, where the sup(sub)script before comma is the index of eigenvectors, and that after comma the index of components in each eigenvector. The same scheme is applied to the vectors $\mathcal{M}^2_{s'_1s'_2s_2}$ and $\mathcal{M}^3_{s'_1s'_2s_2}$. For all of the three types of $\mathcal{M}$ the first and second order corrections to the eigenvalues will go to zero in the thermodynamic limit when the level spacing and the coefficient $B_s$ go to zero. And the corrected eigenvectors are calculated to be 
\begin{eqnarray}\nonumber\label{expansionm}
 \mathcal{M}^{1,1}_{s,s'} &=& \delta_{s,s'}- \sum_D \frac{U^2 B_s B_{s'}T(D)}{(\epsilon_s-\epsilon_{s'})(D-\epsilon_s)} + O(U^3),\\ \nonumber
\mathcal{M}^{1,2}_{s,s'_1s'_2s_2} &=& \frac{-UB_{s'_1s'_2s_2s}}{\epsilon_{s'_1}+\epsilon_{s'_2}-\epsilon_{s_2}-\epsilon_s}+O(U^2), \\  \mathcal{M}^{2,1}_{s'_1s'_2s_2,s} &=& \frac{UB_{s'_1s'_2s_2s}Q_{s'_1s'_2s_2}}{\epsilon_{s'_1}+\epsilon_{s'_2}-\epsilon_{s_2}-\epsilon_s}+O(U^2),
\end{eqnarray}
where $D=\epsilon_{s'_1}+\epsilon_{s'_2}-\epsilon_{s_2}$ is the eigenvalue of $\mathcal{M}^2_{s'_1s'_2s_2}$ and $\mathcal{M}^3_{s'_1s'_2s_2}$, and 
\begin{eqnarray}\label{def_t}\nonumber
T(D) &=& \sum_{s_1' s_2'} Q_{s_1' s_2'(\epsilon_{s_1 '} + \epsilon_{s_2 '}
-D )}B_{s_1'}^2 B_{s_2'}^2 \\
&& \times B^2(\epsilon_{s_1 '} + \epsilon_{s_2 '}-D).
\end{eqnarray}
Here we only list the components of $\mathcal{M}$ which will be used in the calculation of the current to second order of $U$.

Corresponding to the three types of $\mathcal{M}$, the three types of excitation operators can be expressed as
\begin{eqnarray}\nonumber
\hat A^1_s &=& \hat c^\dag_{s\uparrow} + \sum_{s'\neq s } \mathcal{M}^{1,1}_{s,s'} \hat c^\dag_{s'\uparrow}\\ && \nonumber + \sum_{s'_1s'_2s_2} \mathcal{M}^{1,2}_{s,s'_1s'_2s_2}:\hat c^\dag_{s'_1\uparrow} \hat c^\dag_{s'_2\downarrow} \hat c_{s_2\downarrow}: \\ && \nonumber + \mathcal{M}^{1,3}_{s,s'_1s'_2s_2}:\hat c^\dag_{s'_1\uparrow} \hat c^\dag_{s'_2\uparrow} \hat c_{s_2\uparrow}:,
\end{eqnarray} 
\begin{eqnarray}
\nonumber  \hat A^2_{s'_1s'_2s_2} &=& \sum_{s} \mathcal{M}^{2,1}_{s'_1s'_2s_2,s}\hat c^\dag_{s\uparrow}+:\hat c^\dag_{s'_1\uparrow} \hat c^\dag_{s'_2\downarrow} \hat c_{s_2\downarrow}: \\ && \nonumber + \sum_{t'_1t'_2t_2\neq s'_1s'_2s_2} \mathcal{M}^{2,2}_{s'_1s'_2s_2,t'_1t'_2t_2}:\hat c^\dag_{t'_1\uparrow} \hat c^\dag_{t'_2\downarrow} \hat c_{t_2\downarrow}: \\ && \nonumber
+ \sum_{t'_1t'_2t_2} \mathcal{M}^{2,3}_{s'_1s'_2s_2,t'_1t'_2t_2}:\hat c^\dag_{t'_1\uparrow} \hat c^\dag_{t'_2\uparrow} \hat c_{t_2\uparrow}: , 
\end{eqnarray}
and
\begin{eqnarray}
\nonumber  \hat A^3_{s'_1s'_2s_2} &=& \sum_{s} \mathcal{M}^{3,1}_{s'_1s'_2s_2,s}\hat c^\dag_{s\uparrow} \\ && \nonumber + \sum_{t'_1t'_2t_2} \mathcal{M}^{3,2}_{s'_1s'_2s_2,t'_1t'_2t_2}:\hat c^\dag_{t'_1\uparrow} \hat c^\dag_{t'_2\downarrow} \hat c_{t_2\downarrow}:\\ && \nonumber +:\hat c^\dag_{s'_1\uparrow} \hat c^\dag_{s'_2\uparrow} \hat c_{s_2\uparrow}: \\ && \nonumber + \sum_{t'_1t'_2t_2\neq s'_1s'_2s_2} \mathcal{M}^{3,3}_{s'_1s'_2s_2,t'_1t'_2t_2}:\hat c^\dag_{t'_1\uparrow} \hat c^\dag_{t'_2\uparrow} \hat c_{t_2\uparrow}: . \\
\end{eqnarray}
In the matrix form these equations can be expressed as
\begin{eqnarray}
 \left ( \begin{array}{c} \hat A^1_s \\ \hat A^2_{s'_1s'_2s_2} \\ \hat A^3_{s'_1s'_2s_2}\end{array}\right)
= \tilde{\mathcal{M}} \left ( \begin{array}{c} \hat c^\dag_{s'\uparrow} \\ :\hat c^\dag_{t'_1\uparrow} \hat c^\dag_{t'_2\downarrow} \hat c_{t_2\downarrow}: \\ :\hat c^\dag_{t'_1\uparrow} \hat c^\dag_{t'_2\uparrow} \hat c_{t_2\uparrow}: \end{array}\right),
\end{eqnarray}
where $\tilde{\mathcal{M}}$ is a $(N+2N^3)\times (N+2N^3)$ matrix with the elements $\mathcal{M}^{1,1}_{s,s'}$, $\mathcal{M}^{3,3}_{s'_1s'_2s_2,t'_1t'_2t_2}$, etc.. We solve this system of linear equations and get the expression of $\hat c^\dag_{s\uparrow}$, i.e.,
\begin{eqnarray}\nonumber
 \hat c^\dag_{s\uparrow} &=& \sum_{s'} (\tilde{\mathcal{M}}^{-1})^{1,1}_{s,s'} \hat A^1_{s'} + \sum_{s'_1s'_2s_2}  (\tilde{\mathcal{M}}^{-1})^{1,2}_{s,s'_1s'_2s_2} \hat A^2_{s'_1s'_2s_2} \\ &&+ \sum_{s'_1s'_2s_2}  (\tilde{\mathcal{M}}^{-1})^{1,3}_{s,s'_1s'_2s_2} \hat A^3_{s'_1s'_2s_2}.
\end{eqnarray}
We notice that the elements of the matrix $(\tilde{\mathcal{M}}-\textbf{1})$ are all in order $U$, where $\textbf{1}$ is the identity matrix. Then for small $U$ the inverse of $\tilde{\mathcal{M}}$ can be expressed by a Neumann series
\begin{eqnarray}\label{neumannseries}
\tilde{\mathcal{M}}^{-1}=\textbf{1}-(\tilde{\mathcal{M}}-\textbf{1})+(\tilde{\mathcal{M}}-\textbf{1})^2-\cdots.
\end{eqnarray}

The excitation energy of $\hat A^1_{s}$ is $\epsilon_s$, and that of $\hat A^2_{s'_1s'_2s_2}$ and $\hat A^3_{s'_1s'_2s_2}$ is $D=\epsilon_{s'_1}+\epsilon_{s'_2}-\epsilon_{s_2}$. Then the evolution of the hybridization operator can be expressed as
\begin{eqnarray}\nonumber\label{coefficientgamma}
&&  e^{i\hat H t} \hat c^\dag_{s\uparrow} e^{-i\hat H t}= \sum_{s''} \big(\sum_{s'} (\tilde{\mathcal{M}}^{-1})^{1,1}_{s,s'} \tilde{\mathcal{M}}^{1,1}_{s',s''}e^{i\epsilon_{s'}t} \\ && \nonumber + \sum_{s'_1s'_2s_2} (\tilde{\mathcal{M}}^{-1})^{1,2}_{s,s'_1s'_2s_2} \tilde{\mathcal{M}}^{2,1}_{s'_1s'_2s_2,s''} e^{i D t} \\&& \nonumber + \sum_{s'_1s'_2s_2} (\tilde{\mathcal{M}}^{-1})^{1,3}_{s,s'_1s'_2s_2}\tilde{\mathcal{M}}^{3,1}_{s'_1s'_2s_2,s''} e^{i D t} \big)\hat c^\dag_{s''\uparrow}+\cdots. \\
\end{eqnarray}
In this expansion we only keep the lowest order term with single creation operator and neglect the higher order terms, because the higher order terms are normal ordered and will not contribute to the expectation value of the current operator. Substituting Eq.~\ref{neumannseries} and~\ref{expansionm} into Eq.~\ref{coefficientgamma}, we will obtain the coefficient before $\hat c^\dag_{s''\uparrow}$ to second order of $U$. After getting the expression of $e^{i\hat H t} \hat c^\dag_{s\uparrow} e^{-i\hat H t}$, we could directly express the evolution of the current operator according to Eq.~\ref{defcurrentop} as
\begin{eqnarray}
 e^{i\hat H t}\hat I_\uparrow  e^{-i\hat H t} = \sum_{sk} \gamma_s (t)e^{-i\epsilon_k t} \hat c^\dag_ {s \uparrow}\hat c_{k - \uparrow}  + h.c.+\cdots,
\end{eqnarray}
where 
\begin{eqnarray}\label{gamma_s}\nonumber
\gamma_s(t) &=& \frac{iVB_s}{2} e^{i\epsilon_s t} + 
\frac {iVB_s U^2}{2} \sum_{s_1,D} T(D) B_{s_1}^2  \\ && \nonumber \times
\left[ \frac{ e^{iDt}-e^{i\epsilon_s t}}{(\epsilon_s-D)(\epsilon_{s_1}-D)}
+\frac{e^{i\epsilon_s t}- e^{i\epsilon_{s_1} t}}{(\epsilon_s-\epsilon_{s_1})(\epsilon_{s_1}-D)}\right]. \\
\end{eqnarray}
Eq.~\ref{gamma_s} reproduces the result in Ref.~\cite{pei10,fujii}, which indicates that our method is a well-defined perturbative method in this context. However, we would like to mention that by employing the numerical routines as diagonalizing the matrix $\mathcal{H}$ our method has the possibility to go beyond the perturbation theory.

\section{Conclusions}

We develop a method for calculating the real time evolution of an observable operator satisfying the Heisenberg equations of motion. The point of this method is to construct the excitation operators of the Hamiltonian and then decompose the observable operator into the linear combination of the excitation operators. We use this method to calculate the evolution of single creation operator in a toy model Hamiltonian inspired by the Hubbard model and the evolution of the current operator in single impurity Anderson model. We expect to perform our method in a lattice system, e.g. the Hubbard model, in future studies. 

\begin{acknowledgments}
We thank S.~Kehrein, M.~Moeckel and M.~Heyl for valuable discussions.
\end{acknowledgments}

\end{document}